\documentclass[aps,twocolumn,superscriptaddress,noshowpacs,11pt]{revtex4-1}
\usepackage{graphicx}
\usepackage{latexsym}
\usepackage{amssymb}
\usepackage{amsmath}
\usepackage{float}

\begin{document}

\title{Gravitational Lensing of a star by a rotating black hole}
\author{V. I. Dokuchaev}\thanks{e-mail: dokuchaev@inr.ac.ru}
\affiliation{Institute for Nuclear Research, Russian Academy of Sciences, Moscow, 117312 Russia}
\affiliation{National Research Nuclear University MEPhI  \\ (Moscow Engineering Physics Institute), Moscow, 115409 Russia}
\author{N. O. Nazarova}\thanks{e-mail: nazarova.mephi@gmail.com}
\affiliation{National Research Nuclear University MEPhI \\ (Moscow Engineering Physics Institute), Moscow, 115409 Russia}


\begin{abstract}
The gravitational lensing of a finite star moving around a rotating Kerr black hole has been numerically calculated. Calculations for the direct image of the star and for the first and second light echoes have been performed for the star moving with an orbital period of 3.22 h around the supermassive black hole SgrA* at the Galactic Center. Time dependences for the observed star position on the celestial sphere, radiation flux from the star, frequency of detected radiation, major and minor semiaxes of the lensed star image have been calculated and plotted. The detailed observation of such lensing requires a space interferometer such as the Russian Millimetron project.
\end{abstract}

\maketitle

It is expected that the Event Horizon Telescope \cite{Fish16,Lacroix13,Johannsen16,Broderick16,Chael16,Kim16,Roelofs17,Doeleman17} will measure in the next three years the shape of shadow \cite{49,Chandra,51,52,Falcke13,57,Cunha15,Abdujabbarov15,Younsi16} of the supermassive black SgrA* hole at the center of the Galaxy, illuminated by either a hot background gas and stars \cite{53,54,55,56,Yang16} or an accretion disk \cite{58,59,60,61,62,63,Johannsen16b}. A black hole at the center of giant elliptic galaxy M87 is another promising candidate for measuring the shape of black hole shadow. \cite{Broderick09,Broderick11,Doeleman12,Johannsen,48,Broderick15,Lacroix16,Akiyama17}. The existence of black holes in the Universe can be proved for the first time in a direct experiment. This will simultaneously provide an experimental strong field test of not only the general relativity but also many other theories of gravity, e.g., $f(R)$, $C^2$, Galilean, Horndeski, mimetic, and multidimensional (see, e.g., \cite{Frolov09,67,68,Amarilla,66,69,Wei15,Singh16,Giddings16,Mureika,Zakharov16,Amir17,Tsupko17b}).

The next qualitatively new stage of investigations will be a detailed study of the shape of the shadow of the SgrA* black hole, as well as the features of motion of objects around it, e.g., individual stars and compact gas clouds \cite{70,71,65,Dokuch14,DokEr15,FizLab,Nucamendi15,Nucamendi16,CliffWill17a,CliffWill17b,Goddi17}. These experiments require an interferometer with an angular resolution much higher than that of the Event Horizon Telescope. One of such promising interferometer projects with an angular resolution of about one nanoarcsecond is the Russian Millimetron space interferometer project \cite{64}. It is very probable that a compact star or a gas cloud can be located near the SgrA* supermassive star. The comparison of the simulation results of gravitational lensing of the star and/or hot gas clouds near the black hole \cite{263,Zakharov05,264,266,265,Virbhadra09,Doeleman09,Tsupko17a} with the data from a Millimetron-type interferometer soon will make it possible to reliably verify (falsify) the known theories of gravity.

We numerically calculated the gravitational lensing of a finite-size star that moves on a circular orbit in the equatorial plane of the Kerr black hole and is observed by a remote telescope. Trajectories of individual photons emitted by the star were calculated in the geometrical optics approximation. We used the formalism and notation from the pioneering works on gravitational lensing in a strong gravitational field \cite{Polnarev72,CunnBardeen72,CunnBardeen73,Luminet75}. We considered a star (or any other emitting compact object) that moves in the equatorial plane of the black hole, $\theta_s=\pi/2$, in the circular orbit with the radius $r_s=20GM/c^2$, where $M$ is the mass of the black hole,  $G$ is the Newton constant, and $c$ is the speed of light. The direction of the orbital motion of the star coincides with the direction of the rotation of the black hole. In the case of the supermassive black hole SgrA* with the mass  $M=(4.2\pm0.2)\,10^6M_\odot$ and the assumed dimensionless spin (Kerr rotation parameter) $a=0.998$, the orbital velocity of the star is $v\simeq0.22c$ and the orbital period is $T\simeq3.22$\,h. An observer (telescope) is located at a fixed point at the distance $r_0\gg GM/c^2$ from the black hole with the polar angle $\cos\theta_0=0.1$ such that $\theta_0\simeq84^\circ\!\!.24$ and the azimuth angle $\varphi_0=0$. The star is spherical, has the radius $d\ll r_s$, and emits light with a thermal spectrum isotropically from each point of its surface at the total luminosity $L$. The observed energy flux from such a star in the Newtonian limit would be $F_0=L/4\pi r^2_0$. The radial coordinate
in the Figures is given in units of $GM/c^2$.

\begin{figure}
	\begin{center}
		\includegraphics[angle=0,width=0.49\textwidth]{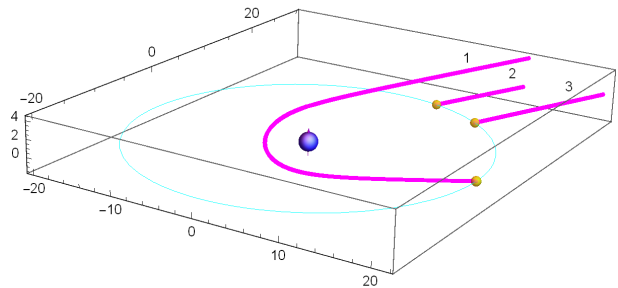}	
	\end{center}
	\caption{$3D$ trajectory of a photon of the first light echo with $\lambda=-6.891,\;q=1.741$ (curve 1), and trajectories of photons of the direct image with $\lambda=-20.95,\;q=2.112\,$ (curve 2), and $\lambda=-0.012,\;q=2.005$  (curve 3).}
	\label{fig1}
\end{figure}

The trajectory of a photon (light geodesic) is independent of its energy and, according to the equations of motion in the Kerr metric \cite{Carter,BPT}, is completely described by two parameters: the dimensionless azimuthal angular momentum $\lambda$ and the dimensionless Carter constant  $q$, which determines the off-equatorial motion of a probe particle. Formally, the remote observer at each time instant sees an infinite number of images of the star. However, the energy flux from most of these images is very low. Figure 1 shows two $3D$ trajectories of photons forming the direct image of the star. These photons fly from the star to the observer, do not intersect the equatorial plane, and always move away from the black hole. Figure 1 also shows the $3D$ trajectory of a photon contributing to the first light echo. These photons fly from the star to observer, first approach the black hole to the minimum distance (turn point) $r_{\rm min}$, and then move away from the black hole, once intersecting the equatorial plane ("one-orbit photons" according to classification in \cite{CunnBardeen73}).  See \cite{Gralla15,Strom16,Gralla16,Strom17,Strom17b} for analytical computations of null geodesics in Kerr space-time. Figure 2 shows a $3D$ trajectory of a photon of the second light echo. These photons fly from the star to observer on trajectories twice intersecting the equatorial plane ("two-orbit photons").
\begin{figure}[h]
	\begin{center}
\includegraphics[angle=0,width=0.45\textwidth]{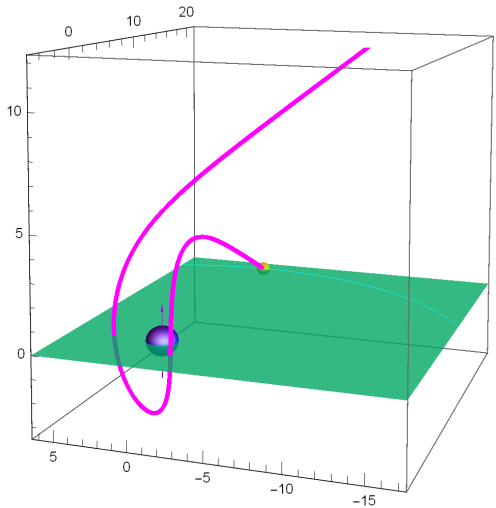}
	\end{center}
\caption{$3D$ trajectory of a photon of the second 	light echo with $\lambda=-1.784,\;q=5.206$.}
	\label{fig2}
\end{figure}

\begin{figure}[h]
	\begin{center}
\includegraphics[width=0.23\textwidth]{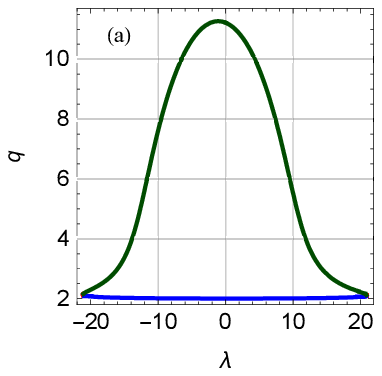} 
		\hfill
\includegraphics[width=0.23\textwidth]{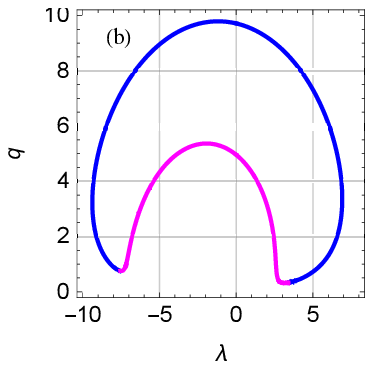} 
		\label{fig3}
	\end{center}
\caption{Parameters of photon trajectories $(\lambda,q)$ forming different types of images of the star: (a) for the direct image, (b) for the first light echo.}
\end{figure}

\begin{figure}[h]
	\begin{center}
\includegraphics[width=0.29\textwidth]{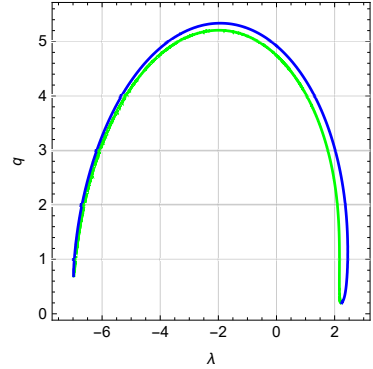}
		\label{fig4}
	\end{center}
\caption{Parameters of photon trajectories $(\lambda,q)$ forming the second light echo.}
\end{figure}

Figures 3 and 4 show the plots of parameters $(\lambda,q)$ for photon trajectories forming the direct image of the star, as well as the first and second light echoes. 

Figure 5 shows the invisible event horizon of the black hole, the shadow of the black hole, the orbit of the direct image of the star, and the orbits of the first and second light echoes. The shadow of the black hole is the limiting internal vicinity of all $N$-th light echoes, where $1\leq N\leq\infty$.

\begin{figure}[h]
	\begin{center}
\includegraphics[angle=0,width=0.49\textwidth]{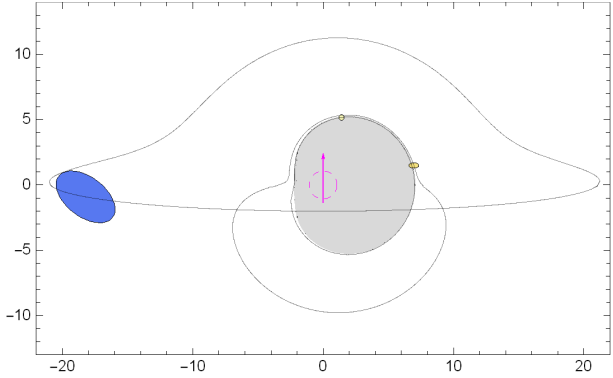}
	\end{center}
\caption{Invisible event horizon of the black hole (dashed circle), rotation axis of the 	black hole (vertical arrow), orbit of the direct image of the star (longest closed curve), orbits of the first and second light echoes adjacent to the shadow of the black hole (filled region), and instantaneous images of the star on these three orbits (filled ellipses).} 
\end{figure}

\begin{figure}[h]
	\begin{center}
\includegraphics[angle=0,width=0.49\textwidth]{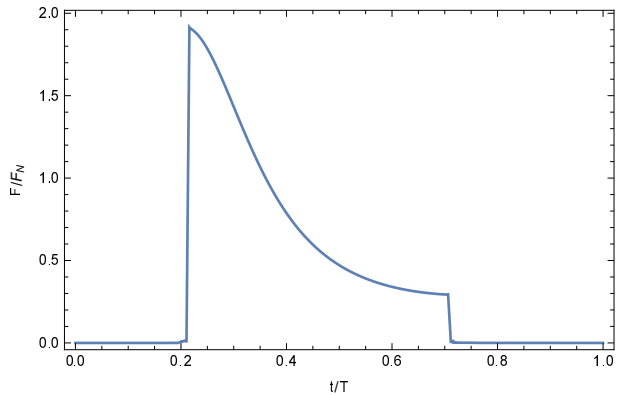}
	\end{center}
\caption{Light curve of the direct image of the star.}
	\label{fig6}
\end{figure}

\begin{figure}[h]
	\begin{center}
\includegraphics[angle=0,width=0.49\textwidth]{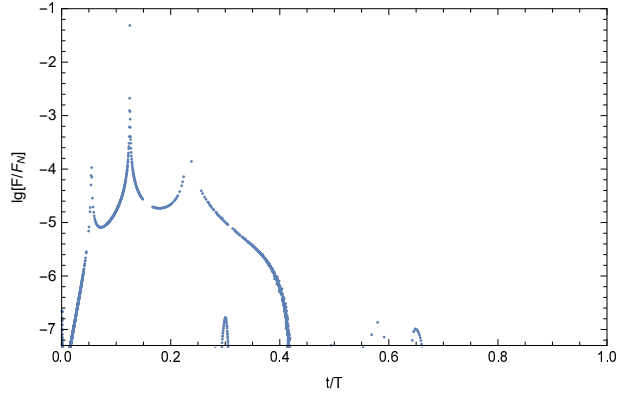}
	\end{center}
\caption{Light curve of the first light echo.}
	\label{fig7}
\end{figure}

\begin{figure}[h]
	\begin{center}
\includegraphics[angle=0,width=0.49\textwidth]{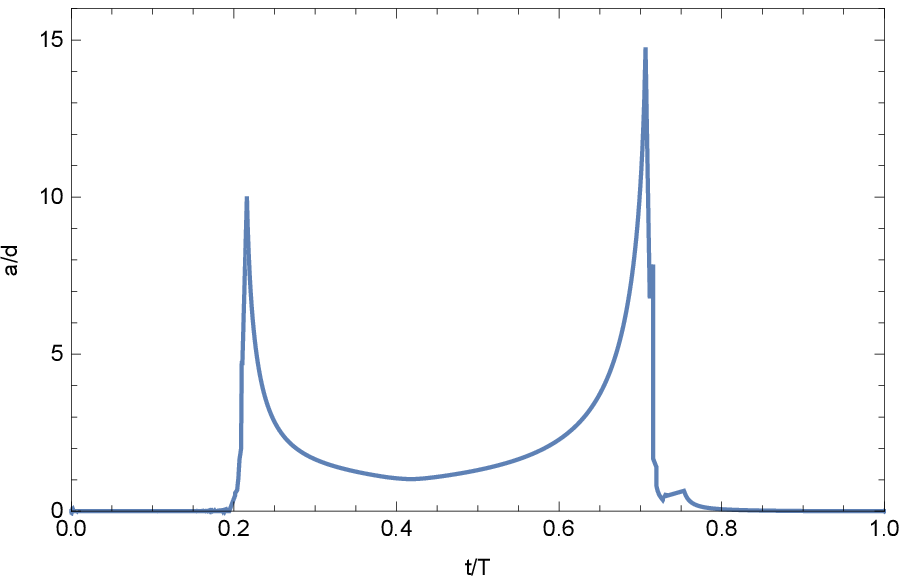}
	\end{center}
\caption{Time dependence of the length of the major semiaxis of the direct image $a(t)$.}
	\label{fig8}
\end{figure}

\begin{figure}[h]
	\begin{center}
\includegraphics[angle=0,width=0.49\textwidth]{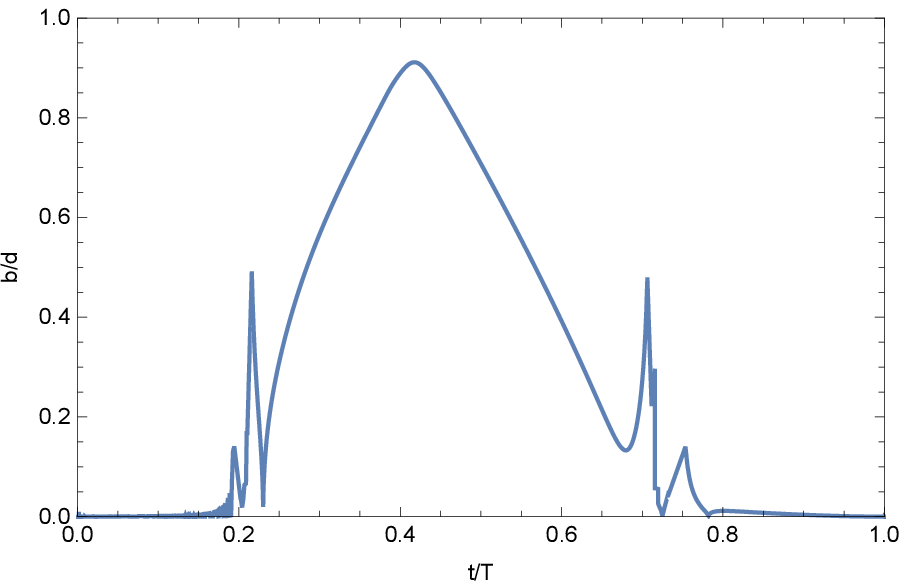}
	\end{center}
\caption{Time dependence of the length of the minor semiaxis of the direct image $b(t)$.}
\end{figure}

\begin{figure}[h]
	\begin{center}
\includegraphics[angle=0,width=0.49\textwidth]{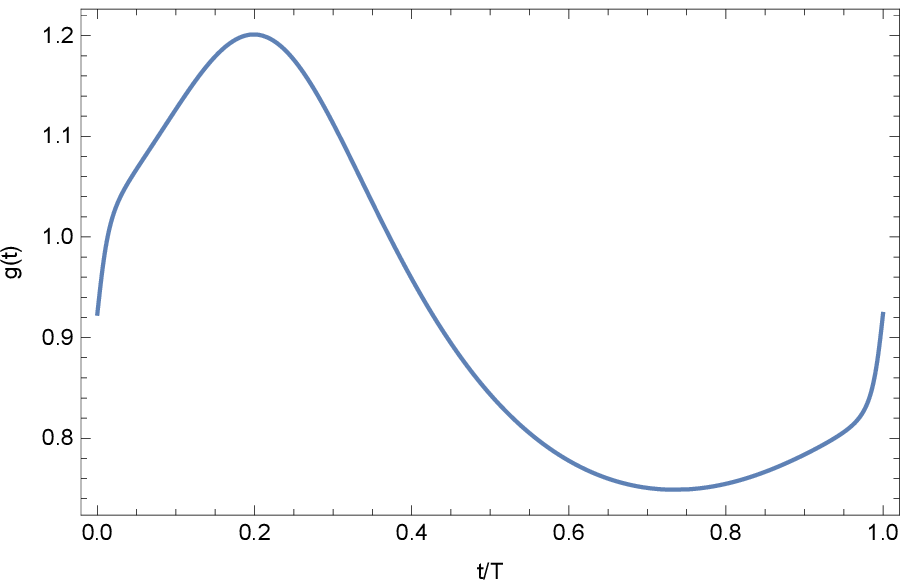}
	\end{center}
\caption{Time dependence of the relative frequency of radiation $g(t)$ of the direct image.}
\end{figure}

When calculating the energy flux detected from various images of the star, we corrected an error in the second integral in Eq. (À20) in \cite{CunnBardeen73}. The correct expression for this integral has the form
\begin{equation}
\int\frac{r[r^3\!-q^2(r\!-\!2)]}{[r^4\!+\!r^2\!+\!2r\!-\!4r\lambda\!
	-\!r(r\!-\!2)\lambda^2\!-\!(r\!-\!1)^2q^2]^{3/2}}dr. \nonumber
\label{power}
\end{equation}

Figures 6 and 7 show light curves for the direct image of the star and the first light echo detected by the remote observer. The time instant $t=0$ on all plots corresponds to the upmost point on the trajectory of the direct image in Fig. 5. The radiation flux $F(t)$ is given in units of the constant Newton radiation flux $F_0$. The radiation flux from the direct image of the star exceeds the Newton value, $F(t)/F_0>1$, on the fragment of the circular orbit, where the star moves toward the observer. This is due to an increase in the visible size of the star and to a strong Doppler effect, which finally results in the compensation of the gravitational redshift. Calculations show that the energy flux from the first light echo is much lower than the energy flux from the direct image on the most part of the orbit. The energy flux from the second light echo (not shown in the figures) is much lower than the energy flux from the first light echo.

\begin{figure}[h]
	\begin{center}
\includegraphics[angle=0,width=0.49\textwidth]{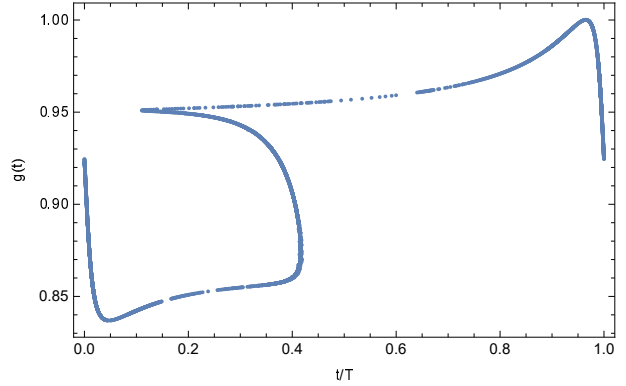}
	\end{center}
\caption{Time dependence of the relative frequency of radiation $g(t)$ of the first light echo. Several images of the first light echo can be observed simultaneously; as a result, the function  $g(t)$ is multivalued.}
\end{figure}

\begin{figure}[h]
	\begin{center}
\includegraphics[angle=0,width=0.49\textwidth]{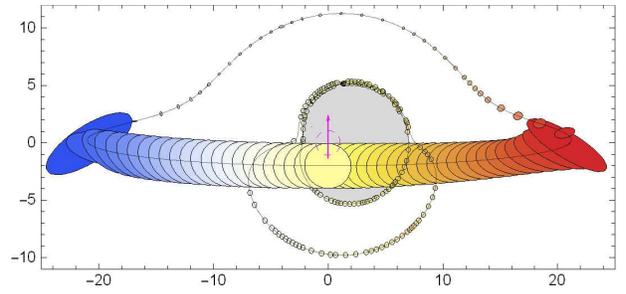}
	\end{center}
\caption{Direct image together with the first and second light echoes through equal time intervals. The size of the second light echo is magnified in order to make it visible.}
\end{figure}

The shape deformation of the lensed images of the spherical star was calculated including the first order of deviation of light rays from the central ray. In this approximation, the instantaneous image of the star is elliptic. The major semiaxis of the ellipse for all images is horizontal in the plane of the sky. Figures 8 and 9 show the time dependences of the lengths of the major  $a(t)$ and minor $b(t)$ semiaxes of the direct elliptic images of the star. We neglect the intrinasic tidal deformation of the star, assuming that the star is quite compact.

Figure 10 shows the time dependence of the relative radiation frequency $g(t)$ from the direct image defined as the ratio of the observed radiation frequency of the star $\nu_0$ to the radiation frequency in the proper reference frame of the star $\nu_s$. Figure 11 shows the corresponding plot of $g(t)$ for the first light echo. The value of $g(t)$ includes both the gravitational redshift in the black hole gravitational field and the Doppler effect caused by the motion of the star. Figure 12 shows direct images of the star and the first and second light echoes at equal time intervals. The time evolution of the discussed lensed images of the star is presented in the animated form in \cite{Youtube}.

In the pioneering paper \cite{CunnBardeen73} on the gravitational lensing in the Kerr metric, the radiation flux from the lensed star was calculated incorrectly because of an error in the integral given by Eq. (1). This error was corrected in our numerical calculations, and the time dependence of radiation flux from the direct image of the star and from the first light echo was recalculated. The numerical algorithm makes it possible to calculate radiation fluxes for any type of light echo. The new result is the calculation of elliptic deformation of the lensed star image in the Kerr metric. These results can be used for interpretation of the future observation data of Millimetron-type projects.

We are grateful to V.~A.~Berezin, Yu.~N.~Ero\-shenko, A.~L.~Smirnov, and A.~F.~Zakharov for stimulating discussions.


\begin{thebibliography}{99}

\bibitem{Fish16} V.\,L. Fish, K. Akiyama, K.\,L. Bouman et al. (Event Horizon Telescope Collaboration), Galaxies {\bf 4}, 54 (2016).

\bibitem{Lacroix13} T. Lacroix and J. Silk, Astron. Astrophys. {\bf 554}, A36 (2013)

\bibitem{Johannsen16} T. Johannsen, A.\,E. Broderick, P.\,M. Plewa, S. Chatzopoulos, S.\,S. Doeleman, F. Eisenhauer, V.\,L. Fish, R. Genzel, O. Gerhard and M.\,D. Johnson, Phys. Rev. Lett. {\bf 116}, 031101 (2016).

\bibitem{Broderick16} A.\,E. Broderick P.\,M. Plewa, S. Chatzopoulos, S.\,S. Doeleman, F. Eisenhauer, V.\,L. Fish, R. Genzel, O. Gerhard and M.\,D. Johnson, Astrophys. J. {\bf 820}, 137 (2016).

\bibitem{Chael16}  A.\,A. Chael, M.\,D. Johnson, R. Narayan, S.\, S. Doeleman, J.\,F.\,C. Wardle and K.\,L. Bouman Astrophys. J. {\bf  829}, 11 (2016).

\bibitem{Kim16} J. Kim, D. P. Marrone, C. Chan, L. Medeiros, F. \"Ozel and D. Psaltis, Astrophys. J. {\bf  832}, 156 (2016).

\bibitem{Roelofs17} F. Roelofs, M. D. Johnson, H. Shiokawa, S. S. Doeleman and H. Falcke, arXiv:1708.01056 [astro-ph.HE].

\bibitem{Doeleman17} S. Doeleman, Nature Astron. {\bf 1}, 646 (2017).

\bibitem{49} J.\,M. Bardeen, in Black Holes (Eds. C. DeWitt, B.\,S. DeWitt) (New York: Gordon and Breach, 1973) p. 215.

\bibitem{Chandra} S. Chandrasekhar, The Mathematical Theory of Black Holes (Oxford: Clarendon Press, 1983).

\bibitem{51} H. Falcke, F. Melia and E. Agol, Astrophys. J. {\bf 528}, L13 (2000).

\bibitem{52} R. Takahashi, Astrophys. J. {\bf 611}, 996 (2004).

\bibitem{Falcke13} H. Falcke and S. Markoff. Class. Quantum Grav. {\bf 30}, 244003 (2013).

\bibitem{57} Z. Li and C. Bambi, JCAP {\bf 01}, 041 (2014).

\bibitem{Cunha15} P.\,V.\,P. Cunha, C.\,A.\,R. Herdeiro, E. Radu and H.\,F. Runarsson, Phys. Rev. Lett. {\bf 115}, 211102 (2015).

\bibitem{Abdujabbarov15} A.\,A. Abdujabbarov, L. Rezzolla and B.\,J. Ahmedov,  Mon. Not. R. Astron. Soc. {\bf 454}, 2423 (2015).

\bibitem{Younsi16} Z. Younsi, A. Zhidenko, L. Rezzolla, R. Konoplya and Y. Mizuno, Phys. Rev. D {\bf 94}, 084025 (2016).

\bibitem{53} R. Takahashi, Publ. Astron. Soc. Jpn. {\bf 57}, 273 (2005).

\bibitem{54} R. Takahashi and K. Watarai,  Mon. Not. R. Astron. Soc. {\bf 374}, 1515 (2007).

\bibitem{55} S. Doeleman, Weintroub, A.\,E.\,E. Rogers et al. Nature {\bf 455}, 78 (2008).

\bibitem{56} F. De Paolis, G. Ingrosso, A.\,A. Nucita, A. Qadir and A.\,F. Zakharov. Gen. Rel. Grav. {\bf 43}, 977 (2011).

\bibitem{Yang16} L. Yang and Z. Li, Intern. J. Mod. Phys D {\bf 25}, 1650026 (2016).

\bibitem{58} P.\,J. Armitage and C.\,S. Reynolds, Mon. Not. R. Astron. Soc. {\bf 341}, 1041 (2003).

\bibitem{59} J. Dexter, E. Agol, P.\,C. Fragile and J.\,C. McKinney, Astrophys. J. {\bf 717}, 1092 (2010).

\bibitem{60} M.\, D. Johnson, V.\,L. Fish, S.\,S. Doeleman, A.\,E. Broderick, J.\,F.\,C. Wardle and D.\,P. Marrone Astrophys. J. {\bf 794}, 150 (2014).

\bibitem{61} O.\,J.\,E. von Tunzelmann, P. Franklin and K.\,S. Thorne, Class. Quantum Grav. {\bf 32}, 065001 (2015).

\bibitem{62} D. Psaltis, R. Narayan, V.\,L. Fish, A.\,E. Broderick, A. Loeb and S.\,S. Doeleman, Astrophys. J. {\bf 798}, 15 (2015).

\bibitem{63} V.\,L. Fish, M.\,D. Johnson, R. Lu, Astrophys. J. {\bf 795}, 134 (2014).

\bibitem{Johannsen16b} T. Johannsen, C. Wang, A.\,E. Broderick, S.\, S. Doeleman, V.\,L. Fish, A. Loeb and D. Psaltis,Phys. Rev. Lett. {\bf 117}, 091101 (2016).

\bibitem{Broderick09} A.\, E. Broderick and A. Loeb, Astrophys. J. {\bf 697} 1164 (2009)

\bibitem{Broderick11} A.\, E. Broderick, A. Loeb and M.\,J. Reid, Astrophys. J. {\bf 735} 57 (2011)

\bibitem{Doeleman12} S.\,S. Doeleman, F.\.V. Fish, D.\,E. Schenck et al. Science, {\bf 338}, 355 (2012).

\bibitem{Johannsen} T. Johannsen, D. Psaltis, S. Gillessen, D.\,P. Marrone, F. Ozel, S,\,S. Doeleman and, V.\,L. Fish, Astrophys. J. {\bf 758}, 30 (2012).

\bibitem{48} M. Inoue, J.\,C. Algaba-Marcos, K. Asada et al. Radio Sci. {\bf 49} 564 (2014).

\bibitem{Broderick15} A.\,E. Broderick, R. Narayan, J. Kormendy, E.\,S. Perlman, M.\,J. Rieke and S.\,S. Doeleman, Astrophys. J. {\bf 805}, 179 (2015).

\bibitem{Lacroix16} T. Lacroix, M. Karami, A.\,E. Broderick, J. Silk and C. Boehm, Phys. Rev. D {\bf 96}, 063008 (2017).

\bibitem{Akiyama17} K. Akiyama, K. Kuramochi, S. Ikeda et al. Astrophys. J. {\bf 838}, 1, (2017).

\bibitem{Frolov09} V.\,P. Frolov, I.\,L. Shapiro,  Phys. Rev. D {\bf 80}, 044034 (2009).

\bibitem{67}  A.\,E. Broderick, A. Loeb and R. Narayan, Astrophys. J. {\bf 701}, 1357 (2009).

\bibitem{68} D. Borka, P. Jovanovic, V.\,B. Jovanovic and A.\,F. Zakharov. JCAP {\bf 11}, 050 (2013).

\bibitem{Amarilla} L. Amarilla and E.\,F. Eiroa, Phys. Rev. D {\bf 87}, 044057 (2013).

\bibitem{66} A.\,F. Zakharov, arXiv:1407.2591.

\bibitem{69} A.\,F. Zakharov, D. Borka, V.\,B. Jovanovic and P. Jovanovic, Adv. Space Res. {\bf 54} 1108 (2014).

\bibitem{Wei15}  S. Wei, P. Cheng, Y. Zhong and X. Zhou, JCAP {\bf 08}, 004 (2015).

\bibitem{Singh16} B.\,P. Singh and S.\,G. Ghosh, arXiv:1707.07125 [gr-qc].

\bibitem{Giddings16} S.\,B. Giddings and D. Psaltis, arXiv:1606.07814 [astro-ph.HE].

\bibitem{Mureika} J.\,R. Mureika and G.\,U. Varieschi, \\ arXiv:1611.00399 [gr-qc].

\bibitem{Zakharov16} A.\,F. Zakharov,  P. Jovanovic, D. Borka and V.\,B. Jovanovic, V. JCAP, {\bf 05}, 045 (2016).

\bibitem{Amir17} M. Amir, B.\,P. Singh and S.\,G. Ghosh, \\ arXiv:1707.09521 [gr-qc]

\bibitem{Tsupko17b} O. Y. Tsupko, Phys. Rev. D  {\bf 95}, 104058 (2017).

\bibitem{70} N.\,S. Kardashev, I.\,D. Novikov and A.\,A. Shatskiy, Int. J. Mod. Phys. D {\bf 16}, 909 (2007).

\bibitem{71} A.\,A. Shatskiy, I.\,D. Novikov and N.\,S. Karda\-shev. Phys. Usp. {\bf 51}, 457 (2008).

\bibitem{65} E.\,O. Babichev, V.\,I. Dokuchaev and Yu.\,N. Ero\-shenko, Phys. Usp. {\bf 56}, 1155 (2013).

\bibitem{Dokuch14} V.\,I. Dokuchaev, Gen. Relativ. Grav. {\bf 46}, 1832 (2014).

\bibitem{DokEr15} V.\,I. Dokuchaev and Yu.\,N. Eroshenko, JETP Lett. {\bf 101}, 777 (2015).

\bibitem{FizLab} V.\,I. Dokuchaev and Yu.\,N. Eroshenko, Phys. Usp, {\bf 58}, 772 (2015).

\bibitem{Nucamendi15} A. Herrera-Aguilar and U. Nucamendi. Phys. Rev. D, {\bf 92}, 045024 (2015).

\bibitem{Nucamendi16} R. Becerril, S. Valdez-Alvarado and U. Nucamendi, Phys. Rev. D {\bf 94}, 124024 (2016).

\bibitem{CliffWill17a} C.\,M. Will and M. Maitra. Phys. Rev. D {\bf 95}, 064003 (2017).

\bibitem{CliffWill17b} F. Ferrer and A.\,M. da Rosa, C.\,M. Will, arXiv:1707.06302 [astro-ph.CO].

\bibitem{Goddi17} C. Goddi, H. Falcke, M. Kramer, et al, Int. J. Mod. Phys. D {\bf 26}, 1730001 (2017).

\bibitem{64} N.\,S. Kardashev, I.\,D. Novikov, V.\,N. Lukash,  et al. Phys. Usp. {\bf 57}, 1199 (2014).

\bibitem{263} A.\,C. Fabian, M.\,J. Rees, L. Stella and N.\,E. White,  Mon. Not. R. Astron. Soc. {\bf 238}, 729 (1989).

\bibitem{Zakharov05} A.\,F. Zakharov, F. De Paolis, G. Ingrosso and A.\,A. Nucita, New Astron. {\bf 10}, 479 (2005).

\bibitem{264} L.\,W. Brenneman and C.\,S. Reynolds, Astrophys J. {\bf 652}, 1028 (2006).

\bibitem{266} A.\,F. Zakharov and S.\,V. Repin, New Astron. {\bf 11}, 405 (2006).

\bibitem{265} J.\,M. Miller, Annu. Rev. Astron. Astrophys. {\bf 45}, 441 (2007).

\bibitem{Virbhadra09} K.\,S. Virbhadra, Phys. Rev. D  {\bf 79}, 083004 (2009)

\bibitem{Doeleman09} S.\,S. Doeleman, V.\,L. Fish,  A.\,E. Broderick, A. Loeb and A.\,E.\,E. Rogers, Astrophys. J.,  {\bf 695}, 59 (2009).

\bibitem{Tsupko17a}  V. Perlick and O.\,Yu. Tsupko, Phys. Rev. D  {\bf 95}, 104003 (2017)

\bibitem{Polnarev72} A.\,G. Polnarev,  Astrofizika {\bf 8}, 273 (1972)].

\bibitem{CunnBardeen72}C.\,T. Cunnungham and J.\,M. Bardeen, Astro\-phys. J. {\bf 173}, L137 (1972).

\bibitem{CunnBardeen73}C.\,T. Cunnungham and J.\,M. Bardeen, Astro\-phys. J. {\bf 183}, 237 (1973).

\bibitem{Luminet75} J.\,-P. Luminet,  Astron. Astrophys. {\bf 75}, 228 (1979).

\bibitem{Carter} B. Carter, Phys. Rev. {\bf 174}, 1559 (1968).

\bibitem{BPT} J.\,M. Bardeen, W.\,H. Press and S.\,A. Teukolsky, Astro\-phys. J. {\bf 178}, 347 (1972).

\bibitem{Gralla15} S. E. Gralla, A. P. Porfyriadis, N. Warburton,
{\it Phys. Rev. D\/} {\bf 92}, 064029 (2015).

\bibitem{Strom16} S. E. Gralla, A. Lupsasca, A. Strominger,
{\it Phys. Rev. D\/} {\bf 93}, 104041  (2016).

\bibitem{Gralla16} S. E. Gralla, A. Zimmerman, P. Zimmerman,
{\it Phys. Rev. D\/} {\bf 94}, 084017 (2016).

\bibitem{Strom17} A. P. Porfyriadis, Y. Shi, A. Strominger,
{\it Phys. Rev. D\/} {\bf 95}, 064009   (2017).

\bibitem{Strom17b}  S. E. Gralla, A. Lupsasca, A. Strominger,
arXiv:1710.11112 [astro-ph.HE].

\bibitem{Youtube} \verb\ https://youtu.be/P6DneV0vk7U \

\end{thebibliography}
\end{document}